\documentclass[pre,twocolumn,showpacs,preprintnumbers,amsmath,amssymb]{revtex4}

\usepackage{graphicx}
\usepackage{dcolumn}
\usepackage{bm}
\usepackage{psfrag}
\usepackage{bbm}

\providecommand*{\dd}{\mathrm{d}}
\providecommand*{\dod}[2]{\frac{\dd #1}{\dd #2}}
\providecommand*{\op}[1]{\hat{#1}}
\providecommand*{\X}{\op\sigma^x}

\providecommand*{\Z}{\op\sigma^z}

\providecommand*{\Hop}{\op H}
\providecommand*{\Tr}{\operatorname{Tr}}
\providecommand*{\Kom}[2]{[#1,#2]}
\providecommand*{\Ket}[1]{\left|#1\right>}
\providecommand*{\Bra}[1]{\left<#1\right|}
\providecommand*{\iu}{\text{i}}

\begin{document}

\preprint{APS/123-QED}

\title{Driven Spin Systems as Quantum Thermodynamic Machines: Fundamental Limits}

\author{Markus J. Henrich}
\email{Markus.Henrich@itp1.uni-stuttgart.de}
\author{G\"unter Mahler}
\affiliation{%
Institute of Theoretical Physics I, University of Stuttgart,
Pfaffenwaldring 57, 70550 Stuttgart, Germany}%
\author{Mathias Michel}
\affiliation{Advanced Technology Institute, School of Electronics and Physical Sciences,
University of Surrey, Gilford, GU2 7XH
United Kingdom}


\begin{abstract}
We show that coupled two level systems like qubits studied in quantum information can be used as a thermodynamic machine. At least three qubits or spins are necessary and arranged in a chain. The system is interfaced between two split baths and the working spin in the middle is externally driven. The machine performs Carnot-type cycles and is able to work as heat pump or engine depending on the temperature difference of the baths $\Delta T$ and the energy differences in the spin system $\Delta E$. It can be shown that the efficiency is a function of $\Delta T$ and $\Delta E$.
\end{abstract}

\maketitle

\section{\label{sec:level1}Introduction}

One of the main goals of thermodynamics has been to study heat engines and thermodynamic processes, dating back to the now famous work of Sardi Carnot in 1824 \cite{Carnot}. With the advent of nano physics and the control of quantum systems down to single atoms a better understanding of thermodynamics on the basis of quantum mechanics is necessary.

Since the first attempts to analyze thermodynamic machines on the quantum level \cite{Scovil1959, Geusic1967} considerable progress has been made in the last decades. Different kinds of models like machines built of harmonic oscillators, of uncoupled spins, particles in a potential or different three level systems have been studied \cite{Feldmann2003, Palao2001, Segal2006, Bender2000, Allahverdyan2005}. 

Also the question about the violation of the second law of thermodynamics in the quantum regime has come up now and then. For a heat engine this would lead to an efficiency larger than the Carnot efficiency. All attempts to do so have failed and could be resolved, e.g., with the help of Maxwell's demon \cite{Kieu2006}. 

Two level systems (TLS) like spins or qubits are the essential ingredients for quantum computation \cite{Loss1998}. Much effort has been directed towards control of small clusters and chains of qubits in quantum optical systems \cite{Cirac2000}, nuclear magnetic resonance \cite{Gershenfeld1997} and solid state systems \cite{Makhlin1999}. A serious problem in any such realization is the interaction of the respective quantum network with its environment.

In the present work we study a model consisting of three TLS arranged in a chain in contact with two baths of different temperatures as studied for transport scenarios, e.g., in \cite{Saito2000,Michel2003,Michel2004}. Here an energy gradient on the system and an incoherent driving of the TLS in the middle let this system act as a thermodynamic machine. For possible experiments the set up may require more TLS's.

Under special conditions the Carnot efficiency may be reached by a TLS heat engine but can never go beyond: If the Carnot efficiency is reached the machine flips its function, e.g., from a heat pump to a heat engine.

We start with a discussion of the concept of work and heat. This is done by considering the change of the energy expection value of a quantum system. With the help of the Gibbs relation heat can be associated with the change of occupation numbers of a quantum system whereas work is the change of the spectrum.

We then introduce our thermodynamic machine consisting of three TLS's \cite{Henrich2006}. Thermodynamic properties can be imparted on this system by an appropriate embedding into a larger quantum environment \cite{GeMiMa2004, Henrich2005, Michel2005}, without the need of any thermal bath. In the present context, though, it is much simpler to settle for the open system approach based on a quantum master equation (QME). In Sec.~\ref{sec:level2} the used QME will be introduced. 

Our numerical results are detailed in Sec.~\ref{sec:level4}. In Sec.~\ref{sec:level5} we compare the numerical investigation with an ideal TLS machine where ideal process steps are assumed. The obtained result is rather general and valid for any kind of TLS machines. 


\section{\label{sec:level23}Thermodynamic Variables}
\subsection{Work and Heat}
To describe thermodynamic processes and machines one first has to define the pertinent variables heat, work, temperature and entropy for the system under consideration. Starting from the energy expectation value
\begin{equation}
 U=\left\langle E \right\rangle = \Tr{\lbrace \Hop \op \rho \rbrace }
\end{equation}
for a quantum system $\Hop$ with discrete spectrum ($\op \rho$ is the density operator) and considering the temporal change of $\left\langle E \right\rangle$
\begin{equation}
 \dod {}{t} \left\langle E \right\rangle = \Tr{\left\lbrace \dod {}{t}\Hop \op \rho\right\rbrace } + \Tr{\left\lbrace \Hop \dod{}{t} \op \rho\right\rbrace },
 \label{eq5}
\end{equation}
change of work $W$ can be associated with the first term of (\ref{eq5}) where only the spectrum changes
\begin{equation}
 \dod {}{t} W=\Tr{\left\lbrace \dod {}{t}\Hop \op \rho\right\rbrace }=\sum_i \dot{E}^i p^i.
 \label{eq6}
\end{equation}
$\dot{E}^i$ is the change per time of the $i$-th eigenvalue and $p^i$ is the corresponding occupation probability. The change of heat $Q$ is then the second part of (\ref{eq5})
\begin{equation}
 \dod {}{t} Q=\Tr{\left\lbrace \Hop \dod {}{t} \op \rho\right\rbrace }=\sum_i \dot{p^i} E^i.
 \label{eq7}
\end{equation}
Equation (\ref{eq5}) thus boils down to the famous Gibbs relation
\begin{equation}
 \Delta U=\Delta W+\Delta Q,
 \label{eq8}
\end{equation}
where $\Delta U$ is the energy change of the system. For cyclic processes work $\Delta W$ and heat $\Delta Q$ can also be calculated with the help of the $S T$-diagram. For a closed path in the $S T$-plane $\Delta U = 0$ and thus
\begin{equation}
  \Delta W = - \Delta Q =- \oint T \dd S.
\label{eq9}
\end{equation}
While connected with bath $\alpha$ $\Delta Q_\alpha$ can alternatively be calculated from the respective heat current $J_\alpha$ over one cycle of duration $\tau$
\begin{equation}
\Delta Q_\alpha=\int_0^\tau J_\alpha \dd t.
\label{eq10}
\end{equation}
Typically there are two baths $\alpha=h,c$ and thus two contributions (see Fig.~\ref{fig1})
\begin{equation}
 \Delta Q= \Delta Q_h + \Delta Q_c.
 \label{eq10b}
\end{equation}

\subsection{Temperature and Entropy}
The temperature of a system can be defined if the state in the energy eigenbasis is canonical. For a TLS $\mu$ the temperature is given by
\begin{equation}
T_\mu=-\frac{E^1_\mu-E^0_\mu}{\text{ln}p^1_\mu-\text{ln}p^0_\mu},
\label{eq11}
\end{equation}
with occupation probability $p_\mu^i$ of the energy level $E_\mu^i$. Due to the fact that all coherences will be damped out by the bath it is always possible to get a local temperature for a single TLS. The von-Neumann-entropy
\begin{equation}
S_\mu=-\Tr{\left\lbrace \op \rho_\mu \ln\op \rho_\mu\right\rbrace }=-\sum_i p^i_\mu \ln p^i_\mu
\label{eq12}
\end{equation}
can then be taken as the thermodynamic entropy.

\subsection{Efficiencies}
The efficiency of a heat pump is defined by the ratio of the heat $\Delta Q_h$ pumped per cycle to the hot reservoir and the work applied 
\begin{equation}
 \eta^p=-\Delta Q_h/\Delta W
 \label{eq14}
\end{equation}
which reduces for the Carnot heat pump to
\begin{equation}
 \eta_\text{Car}^p=T_h/(T_h-T_c).
 \label{eq15}
\end{equation}
 For the heat engine the efficiency is 
 \begin{equation}
  \eta^e=-\Delta W/\Delta Q_h
  \label{eq15a}
 \end{equation}
  which in the Carnot case leads to
 \begin{equation}
  \eta_\text{Car}^e=1-T_c/T_h.
  \label{eq16}
 \end{equation}

\section{Driven Spin System}
\subsection{Hamilton-Model}
\begin{figure}
\centering
\includegraphics[width=.45 \textwidth]{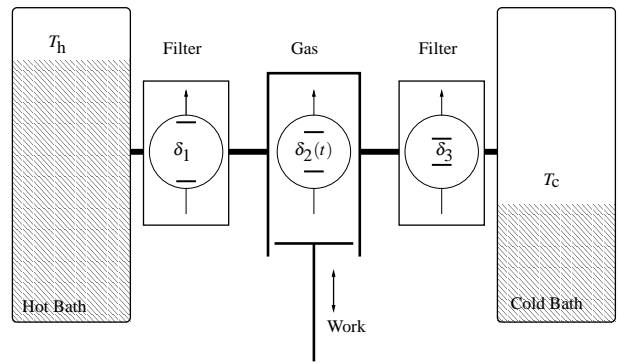}
\caption{Schematic representation of the system under investigation: An inhomogeneous 3-spin-chain is interfaced between two baths. Spin~1 (with energy splitting $\delta_1$) and spin~3 ($\delta_3$) act as filters whereas spin~2 [$\delta_2(t)$] is the working gas by deformation of its spectrum.}
\label{fig1}
\end{figure}
The model under investigation is an inhomogeneous spin chain with nearest neighbor coupling of Heisenberg type described by the Hamiltonian
\begin{equation}
\Hop =\sum_{\mu=1}^3 \left\lbrace  \frac{\delta_\mu}{2} \Z_\mu + \lambda \sum_{i=x,y,z} \op \sigma^i_\mu \otimes \op \sigma_{\mu+1}^i \right\rbrace .
\label{eq1}
\end{equation}
The $\op \sigma^i_\mu$'s are the Pauli-operators of the $\mu$th spin. $\lambda$ is the coupling strength which is chosen to be small compared to the local Zeeman splitting $\delta_\mu$, $\lambda \ll \delta_\mu$. Because $\delta_\mu \ne \delta_{\mu+1}$ we call the spin chain inhomogeneous. 

We will need at least three spins in order to have this system work as a thermodynamic pump or machine. The spin chain is brought in contact locally with two baths at different temperatures as depicted in Fig.~\ref{fig1}. The detuning between spin 1 and spin 3 is $\delta_{13}=(\delta_1-\delta_3)/2>0$.

\subsection{\label{sec:level2}Quantum Master Equation}
There are different ways to describe the thermal behavior of quantum systems coupled to environments. Examples are the path integral method \cite{Weiss} or schemes based on the complete Schr\"odinger dynamics of the small system embedded into a larger quantum environment \cite{GeMiMa2004, Henrich2005, Michel2005}. Because it is much simpler for the present context we settle for a master equation approach. Such an approach has been widely applied to describe system bath models \cite{Breuer,Kubo}.

 To derive the master equation for our model one usually starts from the Liouville-von-Neumann equation for the total system (we set $\hbar$ and the Boltzmann constant $k_B$ equal to 1)
\begin{equation}
 \dod{}{t} \op \rho(t) = -\iu \Kom{\Hop}{\op \rho(t)}.
 \label{eq1y}
\end{equation}
 The Hamiltonian is composed of three terms
  \begin{equation}
   \Hop =\Hop_\text{s}+\Hop_\text{env}+\kappa \Hop_\text{int}
   \label{eq1b}
  \end{equation}
  with $\Hop_\text{s}$ the Hamiltonian of the relevant system, $\Hop_\text{env}$ the environment/bath Hamiltonian, $\Hop_\text{int}$ the interaction of coupling strength $\kappa$ between system and bath.
With the help of a projection operator technique up to second order in $\kappa$ and with the use of the Born-Markov approximation the dynamics of the reduced system density $ \op \rho_\text{s} (t)$ reads
\begin{align}
    & \dod{}{t} \op \rho_\text{s} (t) = \notag \\
    & - \kappa^2 \int_{t_0}^{t} \dd s \Tr_\text{env} \left\lbrace  {\Kom{\Hop_\text{int}(t)}{\Kom{\Hop_\text{int}(t-s)}{\op \rho_\text{s}(t)\otimes \op \rho_\text{env}}}} \right\rbrace ,
   \label{eq1c}
  \end{align}
where $\op \rho_\text{env}$ is a fixed state of the environment and $\text{Tr}_\text{env}$ denotes the trace over all degrees of freedom of the environment (see \cite{Breuer}).

In general the interaction Hamiltonian $\Hop_\text{int}$ is defined as
\begin{equation}
 \Hop_\text{int}=\sum_i \op X_i \otimes \op B_i
 \label{eq1d}
\end{equation}
where $\op X_i$ operates on the system and $\op B_i$ on the environment. For the coupling with a single spin we take $\op X = \X$. By putting (\ref{eq1d}) into (\ref{eq1c}) and going to the Schr\"odinger picture the following compact form can be obtained 
\begin{equation}
 \dod{}{t}\op \rho_\text{s}=-\iu \Kom{\Hop_\text{s}}{\op \rho_\text{s}} + \mathcal{\op D}(\op \rho_\text{s}).
 \label{eq1e}
\end{equation}
As in \cite{Saito2000} we use the dissipator $\mathcal{\op D}(\op \rho_\text{s})$ 
 \begin{equation}
   \op{\mathcal{D}}(\op \rho) = \Kom {\op X}{\op R \op \rho}+\Kom{\op X}{\op R \op\rho}^\dagger\, 
 \label{eq1f}
 \end{equation}
     with
 \begin{equation}
   \Bra{l}\op R \Ket{m}=\Bra{l} \op X \Ket{m} \Phi(E_l-E_m)
   \label{eq1z}
 \end{equation}
 suppressing the system label s in the following. $\Bra{l}$ and $\Ket{m}$ are system eigenstates with the respective energy eigenvalue $E_{l/m}$. $\Phi(E_l-E_m)=\Phi(\omega_{lm})$ is the bath correlation tensor
\begin{equation}
 \Phi(\omega_{lm})=\int_0^\infty \text{e}^{\omega_{lm} s} \langle \op B(s) \op B(0) \rangle \dd s
 \label{eq1g}
\end{equation}
containing the bath correlation function
\begin{equation} 
 \langle \op B(s) \op B(0) \rangle=\Tr_\text{env}{\lbrace \op B(s) \op B(0) \op \rho_\text{env}\rbrace }.
 \label{eq2x}
\end{equation}

Assuming that the state of the bath is a thermal one
 \begin{equation}
  \op \rho_\text{env} = \frac{\text{e}^{- \beta \Hop_\text{env}}}{Z_\text {env}},
  \label{eq2y}
 \end{equation}
($Z_\text{env}$ being the partition function) and that the bath consists of uncoupled harmonic oscillators $\Phi(\omega_{lm})$ takes the form
\begin{equation}
\Phi(\omega_{lm})=\kappa\left( 
\frac{\theta (\omega_{lm})}{\text{e}^{\omega_{lm}\beta_\alpha}-1}+\theta (\omega_{ml})\frac{\text{e}^{\omega_{ml}\beta_\alpha}}{\text{e}^{\omega_{ml}\beta_\alpha}-1}\right).
 \label{eq2e}
\end{equation}
$\theta(\omega_{lm})$ is the step function and $\beta_\alpha$ the respective inverse bath temperature. 

For a three spin chain between two heat baths of different temperatures $T_h$ and $T_c$ and local coupling at the two chain boundaries with
\begin{align}
 \op X_h & = \op \sigma_1^x \otimes \op 1_2 \otimes \op 1_3
 \label{eq2c} \\
 \op X_c & = \op 1_1 \otimes \op 1_2 \otimes \op \sigma^x_3.
 \label{eq2d}
\end{align}
we get instead of (\ref{eq1e}) (cf. \cite{Saito2000})
\begin{equation}
\dod{}{t} \op \rho=-\text{i} \Kom{\Hop}{\op \rho}+\op{\mathcal{D}}_h(\op \rho)+\op{\mathcal{D}}_c(\op \rho).
\label{eq2}
\end{equation}

The stationary state of (\ref{eq1e}) for fixed $\delta_\mu$ is easily shown to be canonical of the form $\op \rho^\text{stat}=\text{e}^{-\beta \Hop_\text{s}}/ \Tr{\{ \text{e}^{-\beta \Hop_\text{s}}\}}$. However, with both baths in place the spin chain might be viewed as a molecular bridge generating a stationary leakage current $J_L=J_h=-J_c$. Here the heat current $J_\alpha$ between the 3-spin-system and the bath $\alpha$ can be defined by the energy dissipated via bath $\alpha$ (cf.~\cite{Breuer})
\begin{equation}
J_\alpha = \Tr{\{ \Hop \op {\mathcal{D}}_\alpha(\op \rho)\} }.
\label{eq3}
\end{equation}
In the following a current out of bath $\alpha$ into the machine will be defined as positive.

\section{\label{sec:level4}Numerical Results}
\subsection{\label{sec:level4a} Non-equilibrium stationary states of a spin chain}
First we note that the heat current through a spin chain depends on the local Zeeman splittings within the system. To analyze the heat current we solve (\ref{eq2}) and calculate the stationary state of the system $\op \rho^\text{stat}$. With this solution and with the help of (\ref{eq3}) we can then calculate the currents $J_\alpha$ for each bath. 

\begin{figure}
\centering
\includegraphics[width=.46 \textwidth]{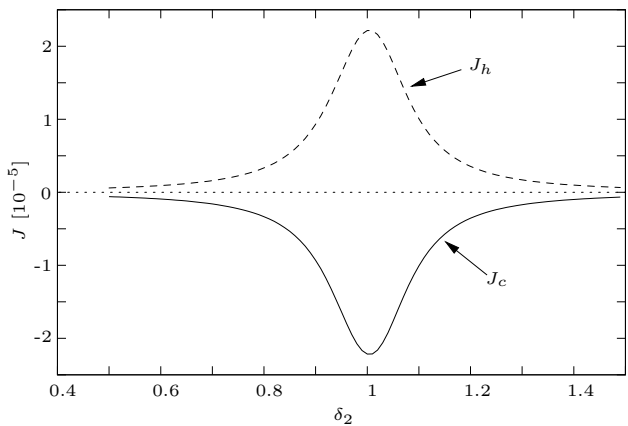}
\caption{Stationary heat current $J_h$ [see (\ref{eq3})] (from hot bath) and $J_c$ (to cold bath) as function of the local energy splitting $\delta_2$ of spin 2 with $\delta_1=\delta_3=1, T_h=2.63$ and $T_c=2.5$.}
\label{fig2}
\end{figure}

We consider a system with $\delta_1=\delta_3=1$. Both heat currents (\ref{eq3}) as function of $\delta_2$ are shown in Fig.~\ref{fig2}. $J_h$ is positive and the relation $J_h = -J_c$ is fullfilled. If $\delta_2=\delta_1=\delta_3$ (the homogeneous case) the heat currents reach their maximum. By detuning the local energy splitting it is thus possible to uncouple the respective bath from the rest of the system. This resonance effect will now be used to build out of three spins a quantum thermodynamic machine.

\subsection{\label{sec:level4b}Time dependent behavior: spin system as thermodynamic heat pump or machine}

\subsubsection{\label{sec:level4b1}The heat current}
The above situation changes when the energy splitting of spin 2 is chosen to be time-dependent, i.e.,
\begin{equation}
 \delta_2(t)=\sin(\omega t) b + b_0.
 \label{eq13}
\end{equation}
Different energy splittings of the boundary spins, e.g.,  $\delta_1=2.25$ and $\delta_3=1.75$, are used to install left/right-selective resonance effects. The parameters of (\ref{eq13}) are chosen as $b_0=(\delta_1+\delta_3)/2$ and $b$ is the detuning $\delta_{13}$. To enable the bath to damp the system $\omega \ll \delta_2$ must be fullfilled. 

For solving (\ref{eq2}) we have used a four step Runge-Kutta-algorithm. At each time step the bath correlation function is calculated explicitly. We choose the following parameters for our numerical results: $\lambda = 0.01$, $\kappa=0.001$, $\delta_1=2.25$, $\delta_3=1.75$, $\omega=1/128$, $T_c=2.5$ and $T_h$ is varied, unless stated otherwise. Both coupling parameters $\lambda$ and $\kappa$ are chosen to stay in the weak coupling limit.

Now when spin 2 is driven periodically as in (\ref{eq13}), we can distinguish four different steps:

\begin{enumerate}
\item Spin 2 (the "working gas") is in resonance with spin 3 [$\delta_2(t) \approx \delta_3$] and thus couples with bath $c$ at temperature $T_c$. Because of this energy resonance the current $J_c$ via spin 3 will be large, whereas the current $J_h$ via spin 2 will be negligible. The occupation probabilities of spin 2 and 3 approach each other and so do the respective local temperatures.

\item Quasi-adiabatic step: Spin 2 is out of resonance with spin 3 [$\delta_1 > \delta_2(t) > \delta_3$], now $J_c$ is suppressed while $J_h$ nearly stays unchanged. The occupation probability of spin 2 does not change significantly and there is almost no change in the entropy $S_2$.

\item Spin 2 is in resonance with spin 1 [$\delta_2(t) \approx \delta_1$] and by that in contact with bath $h$ at temperature $T_h$. $J_h$ is large whereas $J_c$ is very small. The local temperatures of spin 1 and 2 nearly equal each other.

\item Quasi-adiabatic step, as in step 2.
\end{enumerate}

\begin{figure}
\centering%
\includegraphics[width=.46 \textwidth]{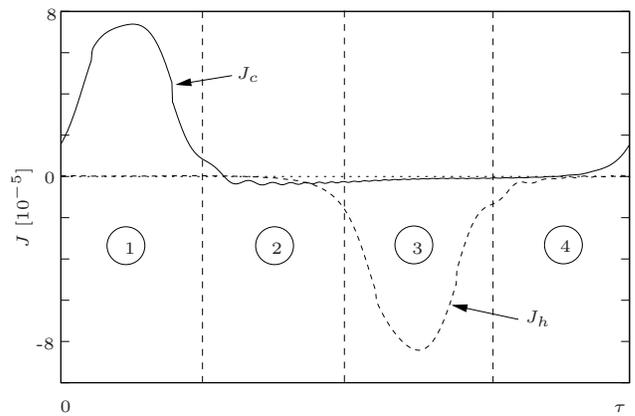}
\caption{Heat currents $J_\alpha(t)$ for the heat pump over one cycle with duration $\tau=2 \pi/\omega=804.25$ for $T_h=2.63$ and $T_c=2.5$. The peaks are a result of the resonance-effect.}
\label{fig3}
\end{figure}

Fig.~\ref{fig3} shows the heat currents $J_\alpha$ of both baths over one period with the bath temperatures $T_h=2.63$ and $T_c=2.5$. As can be seen the resonance effect decouples spin 2 from the bath if its energy splitting is different from the boundary or filter spins. This decoupling is never perfect, though. As a consequence there is a leakage current $J_L$ which will be discussed in more detail later on.

\subsubsection{\label{sec:level4b2}Heat, Work and Efficiencies}
\begin{figure}
\centering%
\includegraphics[width=.46 \textwidth]{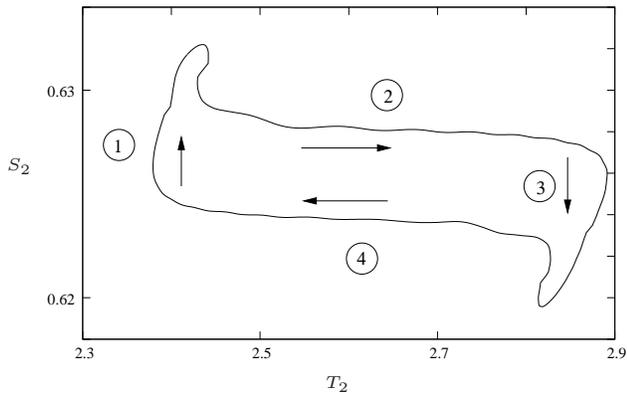}
\caption{$S_2T_2$-Diagram for the quantum heat pump for $T_h=2.63, T_c=2.5$ ($\Delta T=0.13$) and $\tau=2 \pi/\omega=804.25$. The arrows indicate the direction of circulation.}
\label{fig4}
\end{figure}

That the studied system indeed works as a heat pump can be seen from the $S_2T_2$-diagram of spin 2 in Fig.~\ref{fig4}. The local entropy $S_2$ of spin 2 is given by (\ref{eq12}) and the local temperature $T_2$ by (\ref{eq11}). The four different steps as explained in Sec.~\ref{sec:level4b1} are shown as well as the direction of circulation. 

To determine the efficiency of this heat pump one needs to know the quantity of heat $\Delta Q_h$ pumped to the hot bath and the used work $\Delta W$. $\Delta Q_h$ can be calculated by integrating the heat current $J_h$ over one period [cf.~(\ref{eq10})]. 

The exchanged work $\Delta W$ is given by the area enclosed in the $S_2T_2$-plane according to (\ref{eq9}). We find that indeed $\Delta W+\Delta Q_c+\Delta Q_h=0$ in all cases and by that confirm the use of $T_2$ and $S_2$ as effective thermodynamic variables.

\begin{figure}
\centering
 \includegraphics[width=.46 \textwidth]{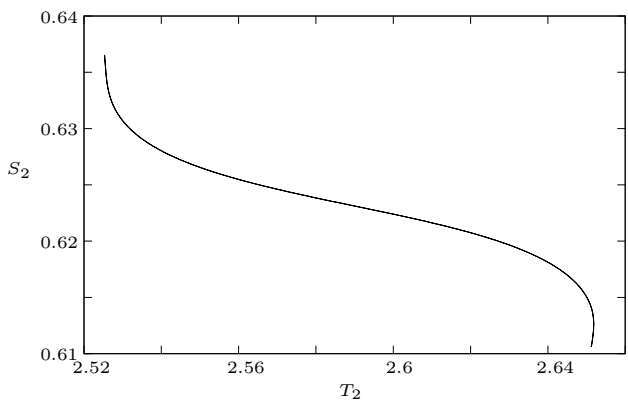}
\caption{$S_2T_2$-diagram for the quasi-statically driven quantum heat pump (with parameters as in Fig.~\ref{fig4}). Because of the leakage current the enclosed $S_2T_2$-area vanishes and no work is exchanged.}
\label{fig7}
\end{figure}

In contrast to the Carnot model our machine is working in finite time. If driven too fast the bath is not able to damp the system and if driven too slow (quasi-stationary) the system would have reached its momentary steady state transport configuration (i.e. $\omega \ll \kappa$). The $S_2T_2$-area then vanishes as depicted in Fig.~\ref{fig7}. This is caused by the leakage current.

\begin{figure}
\centering
 \includegraphics[width=.46 \textwidth]{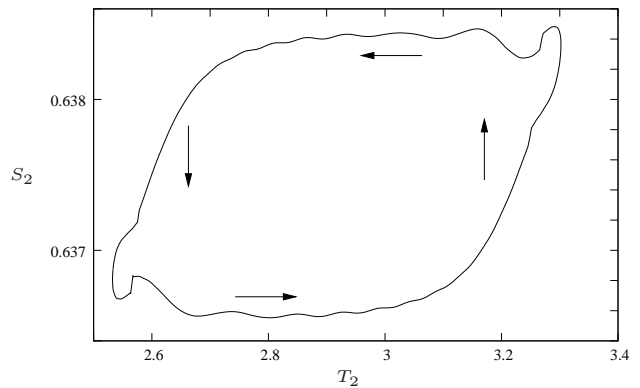}
\caption{$S_2T_2$-Diagram for the quantum heat engine ($\Delta T = 0.83 > \Delta T_\text{max}$ with $T_h=3.33, T_c=2.5$ and $\tau=2 \pi/\omega=804.25$. The arrows indicate the direction of circulation.}
\label{fig11}
\end{figure}

\begin{figure}
\centering
 \includegraphics[width=.47 \textwidth]{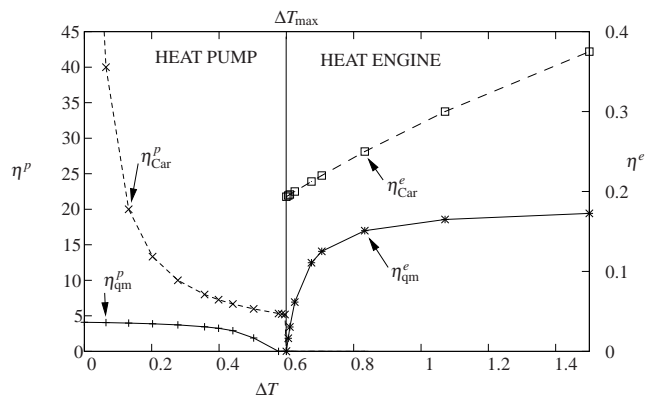}
\caption{The Carnot-efficiency $\eta_\text{Car}^p$ and the efficiency $\eta_\text{qm}^p$ of the quantum heat pump ($\Delta T < \Delta T_\text{max}$) and $\eta_\text{Car}^e$ and $\eta_\text{qm}^e$ of the heat engine ( $\Delta T > \Delta T_\text{max}$) as function of the temperature difference $\Delta T$. Following parameters are chosen: $T_c=2.5, \delta_1=2.25, \delta_3=1.75$ and $\tau=2 \pi/\omega=804.25$.}
\label{fig5}
\end{figure}

Figure~\ref{fig5} shows the Carnot efficiency for the heat pump $\eta_\text{Car}^p$, for the machine $\eta_\text{Car}^e$ and the respective efficiencies for our quantum heat pump $\eta_\text{qm}^p$ [according to (\ref{eq14})] and machine $\eta_\text{qm}^e$ [according to (\ref{eq15a})] as a function of the temperature difference $\Delta T=T_h-T_c$. We point out the following interesting findings:
\begin{itemize}
 \item The efficiency curve of the quantum heat pump or machine is always below the respective Carnot efficiency. As expected, the 2nd law is never violated.
 \item For $\Delta T=0$ $\eta_\text{qm}^p$ does neither diverge nor go to zero. This means that the machine can start out of equilibrium and begin to cool a reservoir.
 \item At a specific temperature difference $\Delta T$, here $\Delta T_\text{max} \approx 0.6$, the heat pump switches to operate as a heat engine. To illustrate this fact Fig.~\ref{fig11} shows the area in the $S_2T_2$-plane for $\Delta T=3.33 > \Delta T_\text{max}$. As depicted, the direction of circulation has reversed.
\end{itemize}

\begin{figure}
\centering
 \includegraphics[width=0.46 \textwidth ]{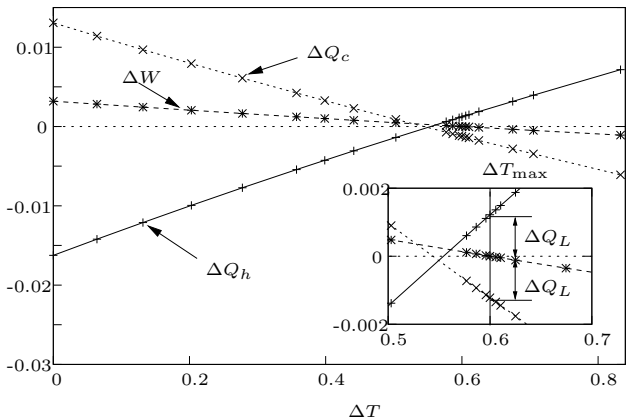}
\caption{Heat $\Delta Q_c$ and $\Delta Q_h$ and work $\Delta W$ performed over one cycle as function of the temperature difference $\Delta T$ (same parameters as in Fig.~\ref{fig5}. The inset shows these functions around the point $\Delta T=\Delta T_\text{max}$ in more detail. $\Delta Q_L$ is the leakage heat per cycle.}
\label{fig6}
\end{figure}

To make the last point more plausible Fig.~\ref{fig6} shows the work $\Delta W$, the heat $Q_h$ and $Q_c$ as function of $\Delta T$. While $\Delta T$ is increasing, $\Delta Q_h$ and $\Delta Q_c$ are decreasing as well as $\Delta W$ until first $\Delta Q_c$ changes its sign, then $Q_h$ and last $\Delta W$. At the point where $\Delta W=0$ (for $\Delta T=\Delta T_\text{max}$) only the leakage current $J_L$ is flowing from the hot bath to the cold one. Beyond this $\Delta T_\text{max}$ the system starts to work as an engine.

\section{\label{sec:level5}Analytical Results}
\subsection{Ideal Quantum Machine}
To understand the above numerical results we compare them to the maximal reachable heat and work which could be pumped or extracted by a TLS quantum machine. All process steps will be taken to be ideal steps. By ideal we mean that we have total control of each process step. Then no leakage current will disturb the system and the heat exchange at bath contact will be without loss. 

In addition we assume a machine which only works during the adiabatic steps. Heat will be exchanged only if the machine is in contact with a bath. This can be compared with the Otto-cycle \cite{Feldmann2003, Feldmann2004}.

We start with spin~2 in contact with spin~3 and thus with cold bath. The state of spin~2 after this contact is a canonical one of the form
\begin{equation} 
 \op \rho_\text{s} = \frac{1}{Z} \left( \begin{array}{cc} \text{e}^{\delta_3/(2T_c)} & 0 \\
    0 &  \text{e}^{-\delta_3/(2T_c)}
                                      \end{array}\right).
 \label{eq50}
\end{equation}
$Z$ is the partition function and we have assumed that the energy of the ground state is $E_2^0=E_3^0=-\frac{\delta_3}{2}$ and the excited state $E_2^1=E_3^1=\frac{\delta_3}{2}$ because both spins are in resonance. 

After this equilibration with the cold bath at $T_c$ the spin~2 is driven until its local energy splitting is equal to spin~1 ($E_2^0=E_1^0=-\frac{\delta_1}{2}$ and $E_2^1=E_1^1=\frac{\delta_1}{2}$). The work for this step can be calculated with (\ref{eq6}). This step is adiabatic as $\op \rho_2$ does not change. The work $W_{3 \rightarrow 1}$ is then given by the energy difference before and after reaching the splitting of spin~1
\begin{equation}
W_{3 \rightarrow 1}=\frac{1}{2} (\delta_3 - \delta_1) \tanh \left( \frac{\delta_3}{2T_c} \right) 
 \label{eq51}
\end{equation}
In contact with spin~1 spin~2 exchanges heat $\Delta Q_h^\text{id}$ with the hot bath at $T_h$. No work will be done and only the occupation probabilities of spin~2 will change to a thermal state with $T_2=T_h$. The exchanged heat can be calculated by the energy difference before  and after thermalisation
\begin{equation}
 \Delta Q_h^\text{id}=\frac{\delta_1}{2} \left[ \tanh \left( \frac{\delta_3}{2T_c}\right) - \tanh \left(\frac{\delta _1}{2T_h} \right)\right] 
 \label{eq52}
\end{equation}
Then spin~2 is driven back to the energy splitting of spin~3 ($E_2^0=E_3^0=\frac{\delta_3}{2}$ and $E_2^1=E_3^1=\frac{\delta_3}{2}$). The work $W_{1 \rightarrow 3}$ for this step is given by
\begin{equation}
W_{1 \rightarrow 3}=\frac{1}{2} (\delta_1 - \delta_3) \tanh \left( \frac{\delta_1}{2T_h} \right).
 \label{53}
\end{equation}
Finally the heat $Q_c^\text{id}$ 
\begin{equation}
 \Delta Q_c^\text{id}=\frac{\delta_3}{2} \left[ \tanh\left( \frac{\delta_1}{2T_h}\right)-\tanh \left(\frac{\delta _3}{2T_c} \right)\right]
 \label{eq54}
\end{equation}
will be exchanged with the cold bath via spin~3. The total work $\Delta W_\text{tot}$ is given by
\begin{equation}
 \Delta W_\text{tot}=W_{3 \rightarrow 1}+W_{1 \rightarrow 3}.
\label{eq54b}
\end{equation}
The Gibbs-relation
\begin{equation}
\Delta W_\text{tot}+\Delta  Q_h^\text{id}+\Delta Q_c^\text{id}=0
\label{eq54c}
\end{equation}
can easily be verified.

With the help (\ref{eq51}) - (\ref{eq54b}) it is now possible to calculated the efficiency of this ideal machine. For the heat pump we get
\begin{equation}
 \eta_\text{id}^p=\frac{-\Delta Q_h^\text{id}}{\Delta W_\text{tot}}=\frac{\delta_1}{\delta_1 - \delta_3},
 \label{eq55}
\end{equation}
for the machine
\begin{equation}
 \eta_\text{id}^e=\frac{-\Delta W_\text{tot}}{\Delta Q_h^\text{id}}=\frac{\delta_1 - \delta_3}{\delta_1}.
 \label{eq56}
\end{equation}
This result is similar to that obtained by Kieu \cite{Kieu2004,Kieu2006} and is the maximum a TLS can reach. Here we want to compare the efficiency of the ideal pump $\eta_\text{id}^p$ and engine $\eta_\text{id}^e$ with the respective Carnot efficiencies for the parameters used for our numerical results.

\begin{figure}[h]
\centering
 \includegraphics[width=0.5 \textwidth ]{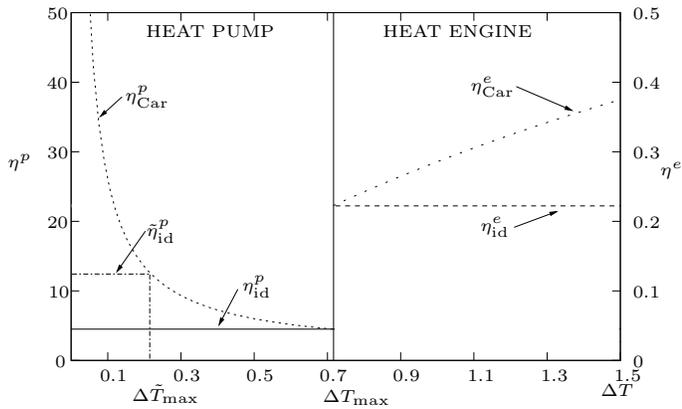}
\caption{Carnot-efficiency $\eta_\text{Car}^p$ for the heat pump and engine $\eta_\text{Car}^e$ as function of temperature difference $\Delta T$ while $T_c=2.5, \delta_1=2.25$ and $\delta_3=1.75$ as in Fig.~\ref{fig5}. $\eta_\text{qm}^p$ and $\eta_\text{qm}^e$ are the efficiencies of the ideal pump/engine [see (\ref{eq55}) and (\ref{eq56})]. $\tilde \eta^p_\text{id}=12.36$ and $\Delta \tilde T_\text{max}=0.22$ can be realized for $\delta_1=1.904$, $\delta_3=1.75$ and $T_c=2.5$.}
\label{fig8}
\end{figure}

Figure~\ref{fig8} shows the Carnot efficiencies as well as the one from (\ref{eq55}) and (\ref{eq56}). $\eta_\text{id}^{p/e}$ is always below  $\eta_\text{Car}^{p/e}$ until it reaches a maximal temperature difference $\Delta T_\text{max}$ (with $T_c=2.5, \delta_1=2.25$ and $\delta_3=1.75$ we get $\Delta T_\text{max}=0.714$). At this temperature the heat pump is working lossless and no heat can be pumped. Just like the quasi-stationary Carnot heat pump this pump has zero power. Only in this particular case $\eta_\text{id}^p =  \eta_\text{Car}^p$. By further increasing the temperature $T_h$ the heat pump starts working as a heat engine. 

\begin{figure}[h]
\centering
 \includegraphics[width=0.46 \textwidth ]{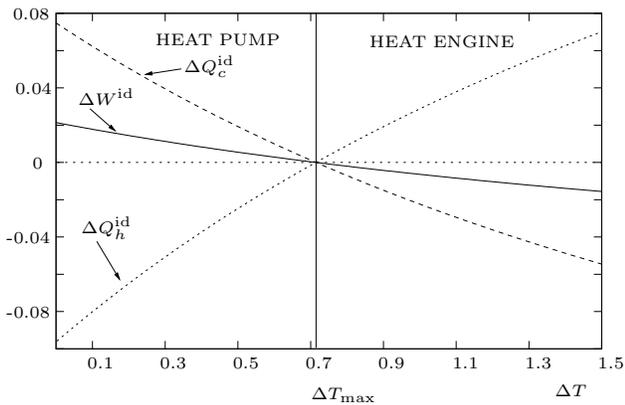}
\caption{Work $\Delta W^\text{id}$, heat $\Delta Q_h^\text{id}$ from/to the hot bath and heat $\Delta Q_c^\text{id}$ from/to the cold bath for the ideal machine as function of temperature difference $\Delta T$ while $T_c=2.5, \delta_1=2.25$ and $\delta_3=1.75$ as in Fig.~\ref{fig6}. At $\Delta T=\Delta T_\text{max}$  $\eta_\text{id}^p =  \eta_\text{Car}^{p/e}$ and therefore $\Delta W^\text{id}=0$, $\Delta Q_h^\text{id}=0$ and $\Delta Q_c^\text{id}=0$.}
\label{fig9}
\end{figure}

Figure~\ref{fig9} illustrates this behavior where $\Delta W^\text{id}, \Delta Q_h^\text{id}$ and $\Delta Q_c^\text{id}$ are depicted as function of $\Delta T$. At $\Delta T_\text{max}$ no heat $\Delta Q_h^\text{id}$ is pumped and therefore no work used or no heat exhausted to do work.

This is qualitatively the same behavior as our model shows in Fig.~\ref{fig5} and Fig.~\ref{fig6}. Two differences can be seen. First the critical temperature in our numerical result deviates from the theoretical expected one. From the numerics we get $\Delta T_\text{max}\approx 0.6$. Second the inset in Fig.~\ref{fig6} shows that $\Delta Q_c$ changes its sign before $\Delta Q_h$ does. The reason for both effects is due to the leakage current as will be explained below.

For a given bath temperature (like in our example $T_c=2.5$) it is possible by changing the energy splittings of $\delta_1$ and/or $\delta_3$ to influence $\Delta T_\text{max}$. In Fig.~\ref{fig8} also a different efficiency $\tilde \eta_\text{id}^p$ is depicted. $\tilde \eta_\text{id}^p$ can be realized by increasing $\delta_1$ so that $\Delta T_\text{max}$ will be decreased to $\Delta \tilde T_\text{max}$.

\subsection{Quantum machine with leakage current}
\begin{figure}
\centering
 \includegraphics[width=0.5 \textwidth ]{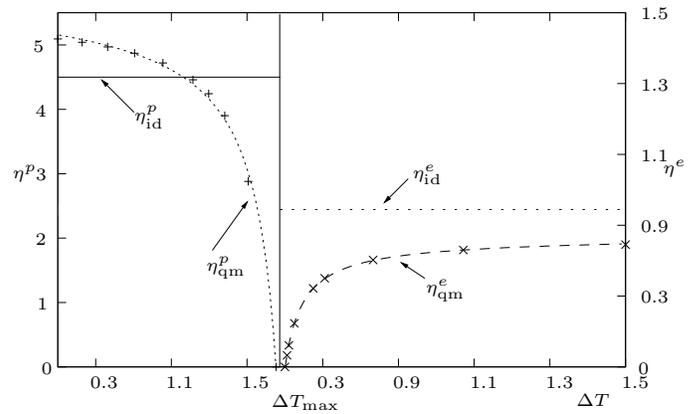}
\caption{Fitted efficiency $\eta_\text{qm}^p$ for the quantum heat pump and quantum engine $\eta_\text{qm}^e$ as function of temperature difference $\Delta T$ while $T_c=2.5, \delta_1=2.25$ and $\delta_3=1.75$ as in Fig.~\ref{fig5}. $\eta_\text{id}^p$ and $\eta_\text{id}^e$ are the efficiencies of the ideal pump/engine [see (\ref{eq55}) and (\ref{eq56})].}
\label{fig10}
\end{figure}

The efficiency of an ideal two level quantum machine is independent of $\Delta T$ except at $\Delta T= \Delta T_\text{max}$, where it jumps between its heat pump and its heat engine value. The efficiency obtained from the numerical simulation deviates somewhat from this expected behavior. For the heat pump the efficiency of our model is even larger than the ideal one (see Fig.~\ref{fig10}). To understand this effect we analyze the leakage current from a phenomenological point of view.

First we assume that the leakage current causes the gas spin~2 to approach a thermal state which is not in accordance with the bath temperature. In this case $\Delta Q_h$ and $\Delta Q_c$ will be decreased. This effect is responsible for the vanishing of $\eta_\text{qm}^{p/m}$ before reaching $\Delta T_\text{max}$. But it can not explain why the efficiency of $\eta_\text{qm}^p$ is sometimes larger than $\eta_\text{id}^p$. 

Taking into account that also less work is performed due to the leakage current it is possible to find a larger efficiency. This can be interpreted in that the gas spin~2 does not ``see'' the full energy splitting $\delta_1$. As shown in Fig.~\ref{fig10} our phenomenological model fits the numerical data quite good.

For the efficiency of the heat engine $\eta_\text{qm}^e$ it can be seen from Fig.~\ref{fig10} that it is always worse than the ideal engine $\eta_\text{id}^p < \eta_\text{qm}^e$.

\section{Conclusion}
We have studied a driven 3-spin system coupled to two split heat baths. We have shown that such small quantum networks may be used not only as quantum information processors but also as quantum thermodynamic machines. For the latter proposal we would primarily exploit the (time-dependent) deformation of discrete spectra and associated resonance transfer. 

While interesting functionality appears already for N = 3 spins, also larger spin networks subject to such very limited control could be envisaged without losing inherent stability: eventually this stability is dictated by the increase of entropy, i.e., by the second law of thermodynamics.

For a thermodynamic TLS machine working with ideal heat transport and adiabatic steps we have derived an ideal efficiency. This efficiency is independent of the bath temperatures. By tuning the energy splitting of the TLS the quantum thermodynamic machine can be used as a heat pump or heat engine. The Carnot efficiency will only be reached when a TLS machine is working losslessly.

Taking dissipation into account it is possible to understand the leakage current present in our numerics from a phenomenological point of view. Surprisingly a leakage current could even increase the efficiency of a heat pump whereas for a heat engine it only decreases the efficiency.

There are a number of different options for implementations \cite{Haefner2005, Maklin2001} and also various possibilities to introduce the time-dependent control. For simplicity we have restricted ourselves here to external driving; alternatively one might look for autonomous system designs \cite{Tonner2005}, e.g., by using a mechanical oscillator (cantilever) \cite{Schwab2005}. Artificial autonomous nanomotors powered by visible light have recently been demonstrated experimentally \cite{Balzani2006}.

From a fundamental point of view several interesting questions remain: What is the status of thermodynamic variables for such quantum systems? To what extent are these measurable in the nano-domain - without being operators? And if measured, how would the measurement result fluctuate \cite{Esposito2006}? 

As noted already, a two-level system diagonal in its local energy basis can always be described as canonical with some temperature $T$, i.e., there is conceptionally no space for nonequilibrium here. It is remarkable that for periodic operation work can then be associated with the area defined by the closed path in the effective entropy-temperature plane for the driven spin, as in macroscopic models. 

This may challenge the subjective ignorance interpretation of none-pure states as classical mixtures, i.e., assuming the individual spin to be either up or down at any time. If the thermal state was taken to result from quantum entanglement with the environment \cite{GeMiMa2004}, this classical picture would no longer be needed; those concepts from quantum information seem to be more appropriate here.

\begin{acknowledgments}
We thank J. Gemmer, F. Rempp, G. Reuther, H. Schmidt, H. Schr\"oder, J. Teifel, P. Vidal and H. Weimer for
fruitful discussions. We thank the Deutsche Forschungsgemeinschaft for
financial support.
\end{acknowledgments}
\bibliographystyle{apsrev}


\end{document}